\documentclass[11pt]{article}
\usepackage[latin9]{inputenc}
\usepackage{amsfonts}
\usepackage{amsmath}
\usepackage{amssymb}
\usepackage{indentfirst}
\usepackage{graphicx}
\usepackage[colorlinks]{hyperref}
\usepackage{cite}

\numberwithin{equation}{section}
\begin{document}

\begin{titlepage}
\vspace{3cm}
\baselineskip=24pt

\begin{center}
\textbf{\LARGE{Non-Relativistic Gravity Theory based on an Enlargement of the Extended Bargmann Algebra}}
\par\end{center}{\LARGE \par}

\begin{center}
	\vspace{1cm}
	\textbf{Patrick Concha}$^{\ast}$,
	\textbf{Evelyn Rodríguez}$^{\dag}$,
	\small
	\\[5mm]
	$^{\ast}$\textit{Instituto
		de Física, Pontificia Universidad Católica de Valparaíso, }\\
	\textit{ Casilla 4059, Valparaiso-Chile.}
	\\[2mm]
	$^{\dag}$\textit{Departamento de Ciencias, Facultad de Artes Liberales,} \\
	\textit{Universidad Adolfo Ibáñez, Viña del Mar-Chile.} \\[5mm]
	\footnotesize
	\texttt{patrick.concha@pucv.cl},
	\texttt{evelyn.rodriguez@edu.uai.cl},
	\par\end{center}
\vskip 20pt
\begin{abstract}
\noindent
In this work we study a non-relativistic three dimensional Chern-Simons gravity theory based on an enlargement of the Extended Bargmann algebra. A finite non-relativistic Chern-Simons gravity action is obtained through the non-relativistic contraction of a particular $U(1)$ enlargement of the so-called AdS-Lorentz algebra. We show that the non-relativistic gravity theory introduced here reproduces the Maxwellian Exotic Bargmann gravity theory when a flat limit $\ell \rightarrow \infty$ is applied. We also present an alternative procedure to obtain the non-relativistic versions of the AdS-Lorentz and Maxwell algebras through the semigroup expansion method.

\end{abstract}
\end{titlepage}\newpage {}

\section{Introduction}

There has been a renewed interest in non-relativistic (NR) theories due to
their utilities to approach strongly coupled condensed matter systems \cite%
{Son, BM, KLM, BG, BGMM, CHOR, CHOR2, Taylor} as well as NR effective field
theories \cite{Son2, HS, GPR, GJA}. At the gravity level, the study of NR
gravitational theories has been discussed by diverse authors in \cite{DK,
DBKP, DGH, Duval, DLP, DLP2, Horava, DH, PS, ABPR, ABGR, BRZ, BR, HLO, BCRR,
CS}. A NR theory can be obtained by a suitable limiting process from a
relativistic theory. In particular, through this work, the NR limit
corresponds to the limit in which the speed of light $c$ tends to infinity.
It is well known that in such NR limit, the AdS spacetime becomes the
Newton-Hooke symmetry which in the flat limit corresponds to the Galilei
symmetry \cite{BLL, BN, ABCP, AMO, Gao, GP, BGK, AGKP}.

On the other hand, three-dimensional gravity models are interesting and
simple toy models as they can be formulated as a Chern-Simons (CS) gauge
theory \cite{Witten, AT, Zanelli}. In particular, CS approach allows us to
construct diverse relativistic and non-relativistic (super)gravity actions.
Nevertheless, the construction of a NR CS action might lead to infinities
and degeneracy. Such difficulties can be overcome by enlarging the field
content of the relativistic theory \cite{BRZ, BR, GO, BCG}. In the case of
the relativistic Poincaré CS gravity theory the NR limit requires to
introduce two central extension in order to admit a non-degenerate bilinear
form. Such extension leads to the exotic Bargmann algebra which allows us to
define a proper finite NR\ CS action \cite{BR, HLO}.

Recently, the NR limit of a three-dimensional CS gravity theory based on a
particular deformation and enlargement of the Poincaré symmetry was
presented in \cite{AFGHZ}. Such symmetry, known as the Maxwell algebra, has
been introduced in \cite{BCR, Schrader, GK} to describe the presence of a
constant electromagnetic field background in a Minkowski space.
Interestingly, the infinities and degeneracy are avoided by considering the
NR contraction of a [Maxwell] $\oplus \,u\left( 1\right) \oplus u\left(
1\right) \oplus u\left( 1\right) $ algebra leading to the Maxwellian Exotic
Bargmann (MEB) algebra.

In this work, we explore the NR\ limit of a three-dimensional CS gravity
theory based on the AdS-Lorentz algebra. Such symmetry is a semi-simple
enlargement of the Poincaré one and has been initially introduced in \cite%
{SS, GKL}. The AdS-Lorentz algebra and its generalizations have been
particularly useful to recover (pure) Lovelock gravity theories from CS and
Born-Infeld theories \cite{CDIMR, CMR, CR3}. At the supersymmetric level,
the supersymmetric extension of the AdS-Lorentz algebra has been used to
introduce alternatively a cosmological constant term in four-dimensional
supergravity \cite{CRS, CIRS, BaRa, PR2}. More recently, a BMS-like ansatz
for a three-dimensional CS theory based on the AdS-Lorentz algebra\ has been
presented in \cite{CMRSV}. In particular, the asymptotic symmetry at null
infinity turns out to be a semi-simple enlargement of the $\mathfrak{bms}%
_{3} $ algebra. An interesting feature of the AdS-Lorentz algebra is that it
reproduces the Maxwell algebra through a flat limit $\ell \rightarrow \infty
$.

Here we present a NR CS gravity based on a particular enlargement of the
extended Bargmann (EEB) algebra by considering the NR contraction of the
[AdS-Lorentz] $\oplus \,u\left( 1\right) \oplus u\left( 1\right) \oplus
u\left( 1\right) $ algebra. Such $U\left( 1\right) $ enlargement not only
allows us to construct a finite NR CS action but also to obtain the MEB
gravity in the flat limit. We also present an alternative procedure to
obtain the EEB and MEB algebras through the semigroup expansion mechanism ($%
S $-expansion) \cite{Sexp, CKMN, AMNT}. Such procedure note only provides us with the
commutators of the NR algebras but also gives the non-vanishing components
of the invariant tensor which are crucial to the proper construction of a
CS\ action.

The paper is organized as follows: in section 2, we give a brief review of
the three-dimensional relativistic AdS-Lorentz CS gravity theory. The
section 3 is devoted to the NR contraction process of the AdS-Lorentz
gravity theory. In section 4, we present an alternative mechanism to obtain
the EEB and MEB algebras by considering the $S$-expansion method. Section 5
is devoted to discussion and possible developments.

Note added: while this manuscript was in the process of typesetting, it came
to our knowledge the ref. \cite{PSR}, which possesses some overlap with
particular cases of our results.

\section{Three-dimensional AdS-Lorentz Chern-Simons gravity, U(1)
enlargements and flat limit}

\subsection{AdS-Lorentz Chern-Simons gravity and flat limit}

In this section, we review the construction of a three-dimensional CS
gravity based on a semi-simple enlargement of the Poincaré group. The
mentioned group is known as the AdS-Lorentz group \cite{SS, DFIMRSV, SaSa}
and the corresponding algebra is a deformation and enlargement of the AdS
algebra. The commutators of the AdS-Lorentz algebra read%
\begin{eqnarray}
\left[ J_{A},J_{B}\right] &=&\epsilon _{ABC}J^{C}\,,\text{ \ \ \ \ }\left[
P_{A},P_{B}\right] =\epsilon _{ABC}Z^{C}\,,  \notag \\
\left[ J_{A},Z_{B}\right] &=&\epsilon _{ABC}Z^{C}\,,\text{ \ \ \ \ }\left[
Z_{A},Z_{B}\right] =\frac{1}{\ell ^{2}}\epsilon _{ABC}Z^{C}\,,
\label{Adslorentz} \\
\left[ J_{A},P_{B}\right] &=&\epsilon _{ABC}P^{C}\,,\text{ \ \ \ \ }\left[
Z_{A},P_{B}\right] =\frac{1}{\ell ^{2}}\epsilon _{ABC}P^{C}\,,  \notag
\end{eqnarray}%
where $J_{A}$ are the spacetime rotations, $P_{A}$ are the spacetime
translations and $Z_{A}$ are non-Abelian generators. Note that $A,B,C$ $%
=0,1,2$ are the Lorentz indices which are lowered and raised with the
Minkowski metric $\eta _{AB}=(-1,1,1)$ and $\epsilon _{ABC}$ is the Levi
Civita tensor which is normalized as $\epsilon _{012}=-\epsilon ^{012}=1.$
Let us note that the name "AdS-Lorentz" is due to the fact that the algebra (%
\ref{Adslorentz}) can be written as the direct sum $\mathfrak{so}\left(
2,2\right) \oplus \mathfrak{so}\left( 2,1\right) $ by a suitable
redefinition of the generators,%
\begin{eqnarray}
J_{A} &=&\ell ^{2}\hat{Z}_{A}\,,  \notag \\
P_{A} &=&\hat{P}_{A}\,, \\
Z_{A} &=&\hat{J}_{A}-\ell ^{2}\hat{Z}_{A}\,.  \notag
\end{eqnarray}

Let us construct now the relativistic CS action for the AdS-Lorentz
symmetry. The three-dimensional CS action is given by%
\begin{equation}
I\left[ A\right] =\int \langle AdA+\frac{2}{3}\,A^{3}\rangle \,,
\label{CSaction}
\end{equation}%
where $A$ is the one-form gauge connection and $\left\langle \cdots
\right\rangle $ denotes the invariant trace. In particular, the one-form $A$
taking values in the AdS-Lorentz algebra (\ref{Adslorentz}) reads%
\begin{equation}
A=W^{A}J_{A}+E^{A}P_{A}+K^{A}Z_{A}\,,  \label{oneform}
\end{equation}%
where $W^{A}$ is the one-form spin connection, $E^{A}$ is the vielbein and $%
K^{A}$ is the gauge field along the non-Abelian generator $Z_{A}\,$. The
associated curvature two-form is%
\begin{equation*}
F=R^{A}(W)J_{A}+R^{A}(E)P_{A}+R^{A}(K)Z_{A}\,,
\end{equation*}%
where%
\begin{eqnarray}
R^{A}(W) &:=&dW^{A}-\frac{1}{2}\epsilon ^{ABC}W_{B}W_{C}\,,  \notag \\
R^{A}(E) &:=&D_{W}E^{A}-\frac{1}{\ell ^{2}}\epsilon ^{ABC}K_{B}E_{C}\,, \\
R^{A}(K) &:=&D_{W}K^{A}-\frac{1}{2\ell ^{2}}\epsilon ^{ABC}K_{B}K_{C}\,-%
\frac{1}{2}\epsilon ^{ABC}E_{B}E_{C}\,.  \notag
\end{eqnarray}%
In particular, the Lorentz covariant derivative is defined as $D_{W}\Theta
^{A}:=d\Theta ^{A}-\epsilon ^{ABC}W_{B}\Theta _{C}$. The non-vanishing
components of an invariant tensor of rank 2 for the AdS-Lorentz algebra are
given by \cite{DFIMRSV, HR, CMRSV}
\begin{eqnarray}
\left\langle J_{A}J_{B}\right\rangle &=&\mu _{0}\eta _{AB}\,,\text{ \ \ \ }%
\left\langle P_{A}P_{B}\right\rangle =\frac{\mu _{2}}{\ell ^{2}}\eta _{AB}\,,
\notag \\
\left\langle J_{A}P_{B}\right\rangle &=&\frac{\mu _{1}}{\ell }\eta _{AB}\,,%
\text{ \ \ }\left\langle Z_{A}Z_{B}\right\rangle =\frac{\mu _{2}}{\ell ^{4}}%
\eta _{AB}\,,  \label{invten} \\
\left\langle J_{A}Z_{B}\right\rangle &=&\frac{\mu _{2}}{\ell ^{2}}\eta
_{AB}\,,\text{ \ \ }\left\langle Z_{A}P_{B}\right\rangle =\frac{\mu _{1}}{%
\ell ^{3}}\eta _{AB}\,,  \notag
\end{eqnarray}%
where the arbitrary constants $\mu _{0}$, $\mu _{1}$ and $\mu _{2}$ can be
redefined as
\begin{equation}
\mu _{0}\rightarrow \alpha _{0}\,,\text{ \ \ \ \ }\mu _{1}\rightarrow \alpha
_{1}\ell \,,\text{ \ \ \ \ }\mu _{2}\rightarrow \alpha _{2}\ell ^{2}\,.
\label{redef}
\end{equation}%
In this way the flat limit $\ell \rightarrow \infty $ is well defined and (%
\ref{invten}) with the redefinition (\ref{redef}) leads to the invariant
tensor of the Maxwell group in three-dimensions. Note that applying this
limit in (\ref{Adslorentz}) leads us to the commutation relations of the
Maxwell algebra. As we will see in the next sections, this behaviour also
arises at the level of the non-relativistic gravities.

Considering the AdS-Lorentz connection one-form (\ref{oneform}) and the
invariant tensor (\ref{invten}) with the redefinition (\ref{redef}) in the
general form (\ref{CSaction}), we find that the explicit form of the
relativistic CS gravity action invariant under the AdS-Lorentz symmetry
reads
\begin{align}
I_{R}=& \int \left[ \alpha _{0}\left( W^{A}dW_{A}+\frac{1}{3}\,\epsilon
^{ABC}W_{A}W_{B}W_{C}\right) \right.  \notag \\
& +\left. \alpha _{1}\left( 2E_{A}R^{A}(W)+\frac{2}{\ell ^{2}}%
\,E_{A}F^{A}(K)+\frac{1}{3\ell ^{2}}\,\epsilon ^{ABC}E_{A}E_{B}E_{C}\right)
\right.  \notag \\
& +\left. \alpha _{2}\left( T^{A}E_{A}+\frac{1}{\ell ^{2}}\,\epsilon
^{ABC}E_{A}K_{B}E_{C}+2K_{A}R^{A}(W)+\frac{2}{\ell ^{2}}K_{A}\,D_{W}K^{A}+%
\frac{1}{3\ell ^{4}}\epsilon ^{ABC}K_{A}K_{B}K_{C}\right) \rule{0pt}{15pt}%
\right] \,\,,  \label{cs}
\end{align}%
where we have defined the torsion two-form $T^{A}:=D_{W}E^{A}$, and the
curvature $F^{A}(K):=D_{W}K^{A}-\frac{1}{2\ell ^{2}}\epsilon
^{ABC}K_{B}K_{C}\,.$ From (\ref{cs}) we see that the action is split into
three different sectors proportional to the $\alpha ^{\prime }$s constants.
The gravitational CS term is related to the $\alpha _{0}$ constant, while
the EH term plus a cosmological constant term appear along the $\alpha _{1}$
constant together with a term depending on the gauge field $K^{A}$. The last
term proportional to $\alpha _{2}$ contain a torsional term plus couplings
of the gravitational gauge fields with the $K^{A}$ field. It is important to
point out that each of the three sectors is invariant under the AdS-Lorentz
symmetry. In particular, the infinitesimal gauge transformations $\delta
_{\Lambda }A=d\Lambda +[A,\Lambda ]$, with gauge parameter $\Lambda =\rho
^{A}J_{A}+\varepsilon ^{A}P_{A}+\gamma ^{A}Z_{A}$, are given by%
\begin{eqnarray}
\delta _{\Lambda }W^{A} &=&D_{W}\rho ^{A}\text{\thinspace }\,,  \notag \\
\delta _{\Lambda }E^{A} &=&D_{W}\varepsilon ^{A}-\epsilon ^{ABC}\rho
_{B}E_{C}-\frac{1}{\ell ^{2}}\epsilon ^{ABC}\left( K_{B}\varepsilon
_{C}-\gamma _{B}E_{C}\right) \,,  \label{gauge} \\
\delta _{\Lambda }K^{A} &=&D_{W}\gamma ^{A}-\epsilon ^{ABC}\left(
E_{B}\varepsilon _{C}-\rho _{B}K_{C}\right) -\frac{1}{\ell ^{2}}\epsilon
^{ABC}K_{B}\gamma _{C}\,.  \notag
\end{eqnarray}%
The CS action (\ref{cs}) is invariant, modulo boundary terms, under these
gauge transformations.

Note that the flat limit $\ell \rightarrow \infty $ applied to (\ref{cs})
leads to the relativistic CS action for the Maxwell symmetry in
three-dimensions \cite{SSV, AFGHZ, CMMRSV}. In the same way, this limit
leads to the gauge transformations for the Maxwell symmetry when is
considered in (\ref{gauge}).

The field equations coming from (\ref{cs}) are given by%
\begin{eqnarray*}
\delta W^{A} &:&\text{ \ \ }0=\alpha _{0}R_{A}(W)+\alpha _{1}\left( T_{A}-%
\frac{1}{\ell ^{2}}\epsilon _{ABC}K^{B}E^{C}\right) +\alpha _{2}\left(
F_{A}(K)-\frac{1}{2}\,\epsilon _{ABC}E^{B}E^{C}\right) \,\,, \\
\delta E^{A} &:&\text{ \ \ }0=\alpha _{1}\left( R_{A}(W)+\frac{1}{\ell ^{2}}%
F_{A}(K)-\frac{1}{2\ell ^{2}}\,\epsilon _{ABC}E^{B}E^{C}\right) +\alpha
_{2}\left( T_{A}-\frac{1}{\ell ^{2}}\epsilon _{ABC}K^{B}E^{C}\,\right) \,, \\
\delta K^{A} &:&\text{ \ \ }0=\frac{\alpha _{1}}{\ell ^{2}}\left( T_{A}-%
\frac{1}{\ell ^{2}}\epsilon _{ABC}K^{B}E^{C}\right) +\alpha _{2}\left(
R_{A}(W)+\frac{1}{\ell ^{2}}F_{A}(K)-\frac{1}{2\ell ^{2}}\,\epsilon
_{ABC}E^{B}E^{C}\right) \,,
\end{eqnarray*}%
which can be written as%
\begin{eqnarray}
R_{A}(W) &=&0\,,  \notag \\
T_{A}-\frac{1}{\ell ^{2}}\epsilon _{ABC}K^{B}E^{C} &=&0\,,  \label{eom2} \\
F_{A}(K)-\frac{1}{2}\,\epsilon _{ABC}E^{B}E^{C} &=&0\,.  \notag
\end{eqnarray}%
As is expected, when $\ell \rightarrow \infty $ in (\ref{eom2}) the
equations lead to the Maxwell field equations.

\ Let us now study the NR limit of the presented theory. Here we are
interested in the limit where the speed of light is taken to infinity. It is
very well-known that taking the limit $c\rightarrow \infty $ in the
relativistic Lagrangian, might lead to infinities. In order to cancel the
divergences, one can include extra fields in the relativistic theory. As we
are also interested in finding the NR Maxwell CS gravity \cite{AFGHZ} in the
limit $\ell \rightarrow \infty $,\ we will consider the [AdS-Lorentz] $%
\oplus \,u(1)\oplus u(1)\oplus u(1)$ theory as our initial relativistic
theory, $i.e$, we add three new extra fields to the field content. In this
way, the NR contraction should lead to a finite Lagrangian and to a NR
algebra with a non-degenerate bilinear form.

Before studying the non-relativistic limit of the AdS-Lorentz gravity
theory, let us first introduce the extra $U(1)$ gauge fields in the one-form
gauge connection (\ref{oneform}).

\subsection{U(1) Enlargements}

Motivated by the discussion of the previous section, we now include three
extra $U(1)$ one-forms gauge fields, $M,$ $S$ and $T$\ in (\ref{oneform}) as
\begin{equation}
A=W^{A}J_{A}+E^{A}P_{A}+K^{A}Z_{A}+MY_{1}+SY_{2}+TY_{3}\,.  \label{A}
\end{equation}%
The new Abelian generators satisfy the following non-zero invariant tensors%
\begin{eqnarray}
\left\langle Y_{1}Y_{1}\right\rangle &=&\alpha _{2}\,,\text{ \ \ \ \ }%
\left\langle Y_{2}Y_{3}\right\rangle =\alpha _{2}\,,  \notag \\
\left\langle Y_{2}Y_{2}\right\rangle &=&\alpha _{0}\,,\text{ \ \ \ \ }%
\left\langle Y_{3}Y_{3}\right\rangle =\frac{\alpha _{2}}{\ell ^{2}},
\label{invten2} \\
\left\langle Y_{1}Y_{2}\right\rangle &=&\alpha _{1}\,,\text{ \ \ \ \ }%
\left\langle Y_{1}Y_{3}\right\rangle =\frac{\alpha _{1}}{\ell ^{2}}.  \notag
\end{eqnarray}%
Then, the non-vanishing components of the invariant tensor for the new
[AdS-Lorentz] $\oplus \,u(1)\oplus u(1)\oplus u(1)$ algebra are given by (%
\ref{invten}) along with (\ref{invten2}). Considering the new enlarged
one-form gauge connection (\ref{A}) and invariant tensor, the relativistic
CS action is written as%
\begin{align}
I_{R}=& \int \left[ \alpha _{0}\left( W^{A}dW_{A}+\frac{1}{3}\,\epsilon
^{ABC}W_{A}W_{B}W_{C}+SdS\right) \right.  \notag \\
& +\left. \alpha _{1}\left( 2E_{A}R^{A}(W)+\frac{2}{\ell ^{2}}%
\,E_{A}F^{A}(K)+\frac{1}{3\ell ^{2}}\,\epsilon ^{ABC}E_{A}E_{B}E_{C}+2MdS+%
\frac{2}{\ell ^{2}}MdT\right) \right.  \notag \\
& +\left. \alpha _{2}\left( T^{A}E_{A}+\frac{1}{\ell ^{2}}\,\epsilon
^{ABC}E_{A}K_{B}E_{C}+2K_{A}R^{A}(W)+\frac{2}{\ell ^{2}}K_{A}\,D_{W}K^{A}%
\right. \rule{0pt}{15pt}\right. \,\,  \notag \\
& +\left. \left. \frac{1}{3\ell ^{4}}\epsilon ^{ABC}K_{A}K_{B}K_{C}+MdM+2SdT+%
\frac{1}{\ell ^{2}}TdT\right) \right] \,.  \label{CS2}
\end{align}%
In the following we will show that the inclusion of these three extra $U(1)$
gauge fields allows to cancel the divergences that appear in the limiting
process. Let us remark that in this work we are interested in the
contraction of the [AdS-Lorentz] $\oplus \,u(1)\oplus u(1)\oplus u(1)$
algebra since, as we will show in the next section, the resulting NR algebra
admits on one hand a non-degenerate invariant tensor, and on the other hand,
leads to the MEB algebra introduced in \cite{AFGHZ} in the $\ell \rightarrow
\infty $ limit.

\section{Non-Relativistic Chern-Simons AdS-Lorentz Gravity}

\subsection{Contraction Process and Enlarged Extended Bargmann Algebra}

In the previous section we have constructed an extended relativistic
AdS-Lorentz algebra. Now, considering the Inönü-Wigner contraction of this
algebra we will obtain a NR version of the AdS-Lorentz algebra. To this
purpose, we will consider a dimensionless parameter $\xi $, and we will
express the relativistic generators as a linear combination of new
generators involving the $\xi $ parameter.

We define the contraction process through the identification of the
relativistic generators defining the [AdS-Lorentz] $\oplus \,u(1)\oplus
u(1)\oplus u(1)$ algebra, with the NR generators (denoted with a tilde) as%
\begin{eqnarray}
J_{0} &=&\frac{\tilde{J}}{2}+\xi ^{2}\tilde{S}\,,\text{\ \ \ \ \ }J_{a}=\xi
\tilde{G}_{a}\,,\text{ \ \ \ \ \ \ }Y_{2}=\frac{\tilde{J}}{2}-\xi ^{2}\tilde{%
S}\,,  \notag \\
P_{0} &=&\frac{\tilde{H}}{2\xi }+\xi \tilde{M}\,,\text{ \ \ \ }P_{a}=\tilde{P%
}_{a}\,,\text{ \ \ \ \ \ \ \ }Y_{1}=\frac{\tilde{H}}{2\xi }-\xi \tilde{M}\,,
\label{con} \\
Z_{0} &=&\frac{\tilde{Z}}{2\xi ^{2}}+\tilde{T}\,,\text{ \ \ \ \ }Z_{a}=\frac{%
\tilde{Z}_{a}}{\xi }\,,\text{ \ \ \ \ \ \ }Y_{3}=\frac{\tilde{Z}}{2\xi ^{2}}-%
\tilde{T}\,.  \notag
\end{eqnarray}%
along with the following scaling%
\begin{equation}
\ell \rightarrow \xi \,\ell \text{\thinspace }.  \label{elle}
\end{equation}%
This redefinition is required in order to have a well-defined limit $\xi
\rightarrow \infty $. \ A particular enlargement of the extended Bargamm
algebra, which we have denoted as EEB algebra, comes from the contraction of
(\ref{Adslorentz}):%
\begin{eqnarray}
\left[ \tilde{G}_{a},\tilde{P}_{b}\right] &=&-\epsilon _{ab}\tilde{M}\,,%
\text{ \ \ \ \ \ \ }\left[ \tilde{G}_{a},\tilde{Z}_{b}\right] =-\epsilon
_{ab}\tilde{T}\,,\text{ \ \ \ \ \ \ }\left[ \tilde{H},\tilde{Z}_{a}\right] =%
\frac{1}{\ell ^{2}}\epsilon _{ab}\tilde{P}_{b}\,,  \notag \\
\left[ \tilde{H},\tilde{G}_{a}\right] &=&\epsilon _{ab}\tilde{P}_{b}\,,\text{
\ \ \ \ \ \ \ \ \ \ }\left[ \tilde{J},\tilde{Z}_{a}\right] =\epsilon _{ab}%
\tilde{Z}_{b}\,,\text{ \ \ \ \ \ \ \ }\left[ \tilde{Z}_{a},\tilde{Z}_{b}%
\right] =-\frac{1}{\ell ^{2}}\epsilon _{ab}\tilde{T}\,,  \notag \\
\left[ \tilde{J},\tilde{P}_{a}\right] &=&\epsilon _{ab}\tilde{P}_{b}\,,\text{
\ \ \ \ \ \ \ \ \ }\left[ \tilde{H},\tilde{P}_{a}\right] =\epsilon _{ab}%
\tilde{Z}_{b}\,,\text{ \ \ \ \ \ \ \ }\left[ \tilde{P}_{a},\tilde{Z}_{b}%
\right] =-\frac{1}{\ell ^{2}}\epsilon _{ab}\tilde{M}\,,  \label{NRAdSLorentz}
\\
\left[ \tilde{J},\tilde{G}_{a}\right] &=&\epsilon _{ab}\tilde{G}_{b}\,,\text{
\ \ \ \ \ \ \ \ }\left[ \tilde{P}_{a},\tilde{P}_{b}\right] =-\epsilon _{ab}%
\tilde{T}\,,\text{ \ \ \ \ \ \ \ }\left[ \tilde{Z},\tilde{P}_{a}\right] =%
\frac{1}{\ell ^{2}}\epsilon _{ab}\tilde{P}_{b}\,,  \notag \\
\left[ \tilde{G}_{a},\tilde{G}_{b}\right] &=&-\epsilon _{ab}\tilde{S}\,,%
\text{ \ \ \ \ \ \ \ \ }\left[ \tilde{Z},\tilde{G}_{a}\right] =\epsilon _{ab}%
\tilde{Z}_{b}\,,\text{ \ \ \ \ \ \ \ \ }\left[ \tilde{Z},\tilde{Z}_{a}\right]
=\frac{1}{\ell ^{2}}\epsilon _{ab}\tilde{Z}_{b}\,.  \notag
\end{eqnarray}%
Here $a=1,2,$ $\epsilon _{ab}\equiv \epsilon _{0ab},$ $\epsilon ^{ab}\equiv
\epsilon ^{0ab}$. Such NR algebra contains three central extensions given by
$\tilde{M}$, $\tilde{S}$ and $\tilde{T}$ which are related to the three
extra $U\left( 1\right) $ generators. Note that when we set $\tilde{Z}=%
\tilde{Z}_{a}=\tilde{T}=0$ the extended Bargmann algebra is recovered. On
the other hand, in the $\ell \rightarrow \infty $ limit the EEB algebra
leads to the MEB algebra \cite{AFGHZ}.

It is interesting to note that the algebra (\ref{NRAdSLorentz}) can be
rewritten as three copies of the so-called Nappi-Witten algebra \cite{NW,
FFSS},
\begin{eqnarray}
\left[ \tilde{J}^{\pm },\tilde{G}_{a}^{\pm }\right] &=&\epsilon _{ab}\tilde{G%
}_{b}^{\pm }\,,\qquad \left[ \tilde{G}_{a}^{\pm },\tilde{G}_{b}^{\pm }\right]
=-\epsilon _{ab}\tilde{S}^{\pm }\,,  \notag \\
\left[ \hat{\tilde{J}},\hat{\tilde{G}}_{a}\right] &=&\epsilon _{ab}\hat{%
\tilde{G}}_{b}\,,\qquad \text{\ \ }\left[ \hat{\tilde{G}}_{a},\hat{\tilde{G}}%
_{b}\right] =-\epsilon _{ab}\hat{\tilde{S}}\,.  \label{3NRL}
\end{eqnarray}%
Such algebra can be viewed as a central extension of the homogeneous part of
the Galilei algebra. Let us note that the three copies of the Nappi-Witten
algebra reproduces the EEB algebra (\ref{NRAdSLorentz}) by considering the
following redefinitions%
\begin{eqnarray}
\tilde{G}_{a} &=&\hat{\tilde{G}}_{a}+\tilde{G}_{a}^{+}+\tilde{G}%
_{a}^{-}\,,\qquad \tilde{P}_{a}=\frac{1}{\ell }\left( \tilde{G}_{a}^{+}-%
\tilde{G}_{a}^{-}\right) \,,\qquad \tilde{Z}_{a}=\frac{1}{\ell ^{2}}\left(
\tilde{G}_{a}^{+}+\tilde{G}_{a}^{-}\right) \,,  \notag \\
\tilde{S} &=&\hat{\tilde{S}}+\tilde{S}^{+}+\tilde{S}^{-}\,,\qquad \ \ \tilde{%
M}=\frac{1}{\ell }\left( \tilde{S}^{+}-\tilde{S}^{-}\right) \,,\qquad \ \
\tilde{T}=\frac{1}{\ell ^{2}}\left( \tilde{S}^{+}+\tilde{S}^{-}\right) \,, \\
\tilde{J} &=&\hat{\tilde{J}}+\tilde{J}^{+}+\tilde{J}^{-}\,,\qquad \ \ \
\tilde{H}=\frac{1}{\ell }\left( \tilde{J}^{+}-\tilde{J}^{-}\right) \,,\qquad
\ \ \tilde{Z}=\frac{1}{\ell ^{2}}\left( \tilde{J}^{+}+\tilde{J}^{-}\right)
\,.  \notag
\end{eqnarray}%
A different redefinition of the generators allows us to rewrite the algebra (%
\ref{NRAdSLorentz}) as the direct sum of the extended Newton-Hooke $\left\{
\bar{\tilde{G}}_{a},\bar{\tilde{P}}_{a},\bar{\tilde{S}},\bar{\tilde{M}},\bar{%
\tilde{J}},\bar{\tilde{H}}\right\} $ and the Nappi-Witten algebra $\left\{
\bar{\tilde{Z}}_{a},\bar{\tilde{T}},\bar{\tilde{Z}}\right\} $. In fact, one
could define%
\begin{eqnarray}
\tilde{G}_{a} &=&\bar{\tilde{G}}_{a}+\bar{\tilde{Z}}_{a}\,,\qquad \tilde{P}%
_{a}=\bar{\tilde{P}}_{a}\,,\qquad \tilde{Z}_{a}=\frac{\bar{\tilde{G}}_{a}}{%
\ell ^{2}}\,,  \notag \\
\tilde{S} &=&\bar{\tilde{S}}+\bar{\tilde{T}}\,,\qquad \tilde{M}=\bar{\tilde{M%
}}\,\,,\qquad \tilde{T}=\frac{\bar{\tilde{S}}}{\ell ^{2}}\,, \\
\tilde{J} &=&\bar{\tilde{J}}+\bar{\tilde{Z}}\,\,,\qquad \tilde{H}=\bar{%
\tilde{H}}\,,\qquad \tilde{Z}=\frac{\bar{\tilde{J}}}{\ell ^{2}}\,,  \notag
\end{eqnarray}%
such that the direct sum of the extended Newton-hooke and the Nappi-Witten
algebra appears,%
\begin{eqnarray}
\left[ \bar{\tilde{G}}_{a},\bar{\tilde{G}}_{b}\right] &=&-\epsilon _{ab}\bar{%
\tilde{S}}\,,\qquad \left[ \bar{\tilde{J}},\bar{\tilde{G}}_{a}\right]
=\epsilon _{ab}\bar{\tilde{G}}_{b}\,,  \notag \\
\left[ \bar{\tilde{G}}_{a},\bar{\tilde{P}}_{b}\right] &=&-\epsilon _{ab}\bar{%
\tilde{M}}\,,\qquad \left[ \bar{\tilde{J}},\bar{\tilde{P}}_{a}\right]
=\epsilon _{ab}\bar{\tilde{P}}_{b}\,,  \notag \\
\left[ \bar{\tilde{P}}_{a},\bar{\tilde{P}}_{b}\right] &=&-\frac{1}{\ell ^{2}}%
\epsilon _{ab}\bar{\tilde{S}}\,,\quad \left[ \bar{\tilde{H}},\bar{\tilde{G}}%
_{a}\right] =\epsilon _{ab}\bar{\tilde{P}}_{b}\,,  \label{NH} \\
\left[ \bar{\tilde{H}},\bar{\tilde{P}}_{a}\right] &=&\frac{1}{\ell ^{2}}%
\epsilon _{ab}\bar{\tilde{G}}_{b}\,,  \notag
\end{eqnarray}%
\begin{equation}
\left[ \bar{\tilde{Z}}_{a},\bar{\tilde{Z}}_{b}\right] =-\epsilon _{ab}\bar{%
\tilde{T}}\,,\qquad \left[ \bar{\tilde{Z}},\bar{\tilde{Z}}_{a}\right]
=\epsilon _{ab}\bar{\tilde{Z}}_{b}\,.  \label{NRL}
\end{equation}%
This is similar to the relativistic AdS-Lorentz algebra which can be written
as the direct sum $\mathfrak{so}\left( 2,2\right) \oplus \mathfrak{so}\left(
2,1\right) $. In particular, the Nappi-Witten algebra is characterized by
the presence of a central extension, denoted by $\bar{\tilde{Z}}$. On the
other hand, the Newton-Hooke symmetries have been recently studied in
three-dimensional spacetime in \cite{AMO, Gao, AGKP}. The extended
Newton-Hooke algebra (\ref{NH}), also known as exotic Newton-Hooke algebra
\cite{AGKP} has two central extensions given by the generators $\bar{\tilde{S%
}}$ and $\bar{\tilde{M}}$. Such algebra is the NR version of the AdS algebra
and appears as a NR contraction of a particular enlargement of the
relativistic AdS algebra.

It is important to remark that, even though the NR algebra (\ref{NH})-(\ref%
{NRL}) looks simpler than (\ref{NRAdSLorentz}), it is the basis $\left\{
\tilde{J},\tilde{G}_{a},\tilde{H},\tilde{P}_{a},\tilde{Z},\tilde{Z}_{a},%
\tilde{S},\tilde{M},\tilde{T}\right\} $ which allows us to make contact with
the Maxwellian Exotic Bargmann gravity through a proper flat limit. In
addition, the relativistic AdS-Lorentz CS gravity theory has been studied
previously in the literature \cite{HR, FISV, CMRSV} in the basis $\left\{
J_{A},P_{A},Z_{A}\right\} $ which satisfy (\ref{Adslorentz}). It is then
natural to construct the CS action related to the EEB algebra in the form (%
\ref{NRAdSLorentz}).

\subsection{Non-Relativistic Chern-Simons action}

Let us consider the explicit construction of a NR CS action based on the EEB
algebra obtained in the previous section.

The non-vanishing components of a non degenerate invariant tensor are
obtained from the contraction (\ref{con}) of the relativistic invariant
tensor (\ref{invten}) with the redefinition (\ref{redef})%
\begin{eqnarray}
\left\langle \tilde{J}\tilde{S}\right\rangle &=&-\tilde{\alpha}_{0}\,,\text{
\ \ \ \ \ \ \ \ \ \ \ \ \ \ \ \ \ \ \ \ \ \ \ \ \ \ }  \notag \\
\left\langle \tilde{G}_{a}\tilde{G}_{b}\right\rangle &=&\tilde{\alpha}%
_{0}\delta _{ab}\,,  \notag \\
\left\langle \tilde{G}_{a}\tilde{P}_{b}\right\rangle &=&\tilde{\alpha}%
_{1}\delta _{ab}\,,  \label{inv3a} \\
\left\langle \tilde{H}\tilde{S}\right\rangle &=&\left\langle \tilde{M}\tilde{%
J}\right\rangle =-\tilde{\alpha}_{1}\,,  \notag \\
\left\langle \tilde{P}_{a}\tilde{P}_{b}\right\rangle &=&\left\langle \tilde{G%
}_{a}\tilde{Z}_{b}\right\rangle =\tilde{\alpha}_{2}\delta _{ab}\,,  \notag \\
\left\langle \tilde{J}\tilde{T}\right\rangle &=&\left\langle \tilde{H}\tilde{%
M}\right\rangle =\left\langle \tilde{S}\tilde{Z}\right\rangle =-\tilde{\alpha%
}_{2}\,,  \notag
\end{eqnarray}%
\begin{eqnarray}
\left\langle \tilde{Z}_{a}\tilde{Z}_{b}\right\rangle &=&\frac{\tilde{\alpha}%
_{2}}{\ell ^{2}}\delta _{ab}\,,  \notag \\
\left\langle \tilde{Z}_{a}\tilde{P}_{b}\right\rangle &=&\frac{\tilde{\alpha}%
_{1}}{\ell ^{2}}\delta _{ab}\,,  \label{inv3b} \\
\left\langle \tilde{Z}\tilde{M}\right\rangle &=&\left\langle \tilde{T}\tilde{%
H}\right\rangle =-\frac{\tilde{\alpha}_{1}}{\ell ^{2}}\delta _{ab}\,,  \notag
\\
\left\langle \tilde{Z}\tilde{T}\right\rangle &=&-\frac{\tilde{\alpha}_{2}}{%
\ell ^{2}}\delta _{ab}\,,  \notag
\end{eqnarray}%
where the relativistic parameters $\alpha $'s were rescaled as
\begin{equation}
\alpha _{0}=\tilde{\alpha}_{0}\xi ^{2}\,,\text{ \ \ \ \ }\alpha _{1}=\tilde{%
\alpha}_{1}\xi \,,\text{ \ \ \ \ }\alpha _{2}=\tilde{\alpha}_{2}\,,
\label{alphas}
\end{equation}%
in order to have a finite CS action. As is expected, the flat limit $\ell
\rightarrow \infty $ applied to (\ref{inv3a})-(\ref{inv3b}) leads to the NR
invariant tensors for the MEB algebra \cite{AFGHZ}.

Now we can write the NR one-form gauge connection $\tilde{A}$ in terms of
the NR generators as follows%
\begin{equation}
\tilde{A}=\tau \tilde{H}+e^{a}\tilde{P}_{a}+\omega \tilde{J}+\omega ^{a}%
\tilde{G}_{a}+k\tilde{Z}+k^{a}\tilde{Z}_{a}+m\tilde{M}+s\tilde{S}+t\tilde{T}%
\,.
\end{equation}%
The NR curvature two-form is then%
\begin{eqnarray}
\tilde{F} &=&R(\tau )\tilde{H}+R^{a}(e^{b})\tilde{P}_{a}+R(\omega )\tilde{J}%
+R^{a}(\omega ^{b})\tilde{G}_{a}+R(k)\tilde{Z}  \notag \\
&&+R^{a}(k^{b})\tilde{Z}_{a}+R(m)\tilde{M}+R(s)\tilde{S}+R(t)\tilde{T}\,,
\end{eqnarray}%
where%
\begin{eqnarray}
R(\tau ) &=&d\tau \,,  \notag \\
R^{a}(e^{b}) &=&de^{a}+\epsilon ^{ac}\omega e_{c}+\epsilon ^{ac}\tau \omega
_{c}+\frac{1}{\ell ^{2}}\epsilon ^{ac}ke_{c}+\frac{1}{\ell ^{2}}\epsilon
^{ac}\tau k_{c}\,,  \notag \\
R(\omega ) &=&d\omega \,,  \notag \\
R^{a}(\omega ^{b}) &=&d\omega ^{a}+\epsilon ^{ac}\omega \omega _{c}\,,
\notag \\
R(k) &=&dk\,,  \label{curv} \\
R^{a}(k^{b}) &=&dk^{a}+\epsilon ^{ac}\omega k_{c}+\epsilon ^{ac}\tau
e_{c}+\epsilon ^{ac}k\omega _{c}+\frac{1}{\ell ^{2}}\epsilon ^{ac}kk_{c}\,,
\notag \\
R(m) &=&dm+\epsilon ^{ac}e_{a}\omega _{c}+\frac{1}{\ell ^{2}}\epsilon
^{ac}e_{a}k_{c}\,,  \notag \\
R(s) &=&ds+\frac{1}{2}\epsilon ^{ac}\omega _{a}\omega _{c}\,,  \notag \\
R(t) &=&dt+\epsilon ^{ac}\omega _{a}k_{c}+\frac{1}{2}\epsilon
^{ac}e_{a}e_{c}+\frac{1}{2\ell ^{2}}\epsilon ^{ac}k_{a}k_{c}\,.  \notag
\end{eqnarray}%
Finally, the NR CS action invariant under the EEB algebra is%
\begin{eqnarray}
I_{NR} &=&\int \tilde{\alpha}_{0}\left[ \omega _{a}R^{a}(\omega
^{b})-2sR\left( \omega \right) \right] +\tilde{\alpha}_{1}\left[
2e_{a}R^{a}(\omega ^{b})-2mR(\omega )-2\tau R(s)+\frac{2}{\ell ^{2}}%
e_{a}F^{a}(k^{b})\right.  \notag \\
&&\left. -\frac{2}{\ell ^{2}}\tau R(t)-\frac{2}{\ell ^{2}}mR\left( k\right) %
\right] +\tilde{\alpha}_{2}\left[ e_{a}R^{a}\left( e^{b}\right)
+k_{a}R^{a}\left( \omega ^{b}\right) +\frac{1}{\ell ^{2}}k_{a}R^{a}\left(
k^{b}\right) \right.  \label{NRaction} \\
&&\left. +\omega _{a}R^{a}\left( k^{b}\right) -2sR\left( k\right) -2mR\left(
\tau \right) -2tR\left( \omega \right) -\frac{1}{\ell ^{2}}2tR\left(
k\right) \right] \,,  \notag
\end{eqnarray}%
where%
\begin{equation*}
F^{a}(k^{b}):=dk^{a}+\epsilon ^{ac}\omega k_{c}+\epsilon ^{ac}k\omega _{c}+%
\frac{1}{\ell ^{2}}\epsilon ^{ac}kk_{c}\,.
\end{equation*}%
The NR CS action contains three sectors proportional to different arbitrary
constants $\tilde{\alpha}_{i}$. The first term is the CS action for the NR
Exotic Gravity. The second and third term reproduce the enlarged extended
Bargmann gravity with the explicit presence of the $k_{a}$ gauge field. Let
us note that the limit $\ell \rightarrow \infty $ in the term proportional
to $\tilde{\alpha}_{1}$ reproduces the CS action for the extended Bargmann
algebra \cite{BR, HLO}. On the other hand, the flat limit $\ell \rightarrow
\infty $ in the $\tilde{\alpha}_{2}$ term leads us to the CS action for the
NR\ Maxwell algebra \cite{AFGHZ}. It is important to point out that the term
proportional to $\tilde{\alpha}_{1}$ is not the exotic Newton-Hooke gravity
Lagrangian although it leads to the extended Bargmann gravity Lagrangian in
the $\ell \rightarrow \infty $ limit. In particular, the additional gauge
fields related to the EEB algebra appearing in the $\tilde{\alpha}_{1}$ term
vanish in the flat limit.

Note that we can express the relativistic gauge fields in terms of the NR
ones as follows%
\begin{eqnarray}
W^{0} &=&\omega +\frac{s}{2\xi ^{2}}\,,\text{\ \ \ \ \ }W^{a}=\frac{\omega
^{a}}{\xi }\,,\text{ \ \ \ \ \ \ }S=\omega -\frac{s}{2\xi ^{2}}\,,  \notag \\
E^{0} &=&\xi \tau +\frac{m}{2\xi }\,,\text{ \ \ \ \ }E^{a}=e^{a}\,,\text{ \
\ \ \ \ \ \ \ }M=\xi \tau -\frac{m}{2\xi }\,,  \label{NRfields} \\
K^{0} &=&\xi ^{2}k+\frac{t}{2}\,,\text{ \ \ \ \ }K^{a}=\xi k^{a}\,,\text{ \
\ \ \ \ \ }T=\xi ^{2}k-\frac{t}{2}\,,  \notag
\end{eqnarray}%
such that $A=\tilde{A}$. It is straightforward to show that by using (\ref%
{elle}), (\ref{alphas}) and (\ref{NRfields}) in the action (\ref{CS2}),
after taking the limit $\xi \rightarrow \infty $ the NR action (\ref%
{NRaction}) is obtained.

One can notice that the equations of motion from the finite NR action (\ref%
{NRaction}) reduce to the vanishing of the curvatures (\ref{curv}). A
completely different situation would occur if the invariant tensor became
degenerate during the NR contraction procedure.

\section{Non-Relativistic algebras and S-expansion}

In this section we show that the Enlarged Extended Bargmann algebra (\ref%
{NRAdSLorentz}) can be alternatively obtained using the $S$-expansion
procedure \cite{Sexp}. We show first that the Maxwellian Exotic Bargmann
algebra\textit{\ }can also be obtained using the $S$-expansion method. In
simple words, this method consists in finding a new Lie algebra $\mathfrak{G}%
=S\times \mathfrak{g}$, by combining the structure constants of a Lie
algebra $\mathfrak{g}$ with the elements of a semigroup $S$. Our starting
point will be the so-called Nappi-Witten algebra \cite{NW, FFSS} which can
be interpreted as a central extension of the homogeneous part of the Galilei
algebra. Such Lie algebra can be obtained as a Inönü-Wigner (IW) contraction
of a central extension of the Lorentz algebra. In this way, we consider the
algebra generated by the Lorentz generator $J_{A}$, and one extra $U(1)$
generator $Y$.

The contraction process is defined by the identification of the generators
defining the [Lorentz] $\oplus \,u(1)$ algebra, with the Nappi-Witten
generators as%
\begin{equation}
J_{0}=\frac{\tilde{J}}{2}+\xi ^{2}\tilde{S}\,,\ \ \ \ \ J_{a}=\xi \tilde{G}%
_{a}\,,\ \ \ \ \ \ Y=\frac{\tilde{J}}{2}-\xi ^{2}\tilde{S}\,,
\end{equation}%
where $\tilde{J}$ are spatial rotations, $\tilde{G}_{a}$ are Galilean boosts
and $\tilde{S}$ is a central charge. The resulting algebra coming from the
contraction of the central extension of the Lorentz algebra is%
\begin{eqnarray}
\left[ \tilde{J},\tilde{G}_{a}\right] &=&\epsilon _{ab}\tilde{G}_{b}\,,
\notag \\
\left[ \tilde{G}_{a},\tilde{G}_{b}\right] &=&-\epsilon _{ab}\tilde{S}\,.
\label{NRLorentz}
\end{eqnarray}%
This algebra, which can be viewed as the Heisenberg algebra with an outher
automorphism $J$, will be our original algebra in the expansion process. As
we will see, using different Abelian semigroups it is possible to derive
different NR algebras.

\subsection{Maxwellian Exotic Bargmann Algebra}

Let us first obtain the Maxwellian Exotic Bargmann algebra recently
introduced in \cite{AFGHZ}. For this purpose we consider the Nappi-Witten
algebra (\ref{NRLorentz}) and the Abelian semigroup $S_{E}^{(2)}=\left\{
\lambda _{0},\lambda _{1},\lambda _{2},\lambda _{3}\right\} $, whose
multiplication law is given by%
\begin{equation}
\begin{tabular}{l|llll}
$\lambda _{3}$ & $\lambda _{3}$ & $\lambda _{3}$ & $\lambda _{3}$ & $\lambda
_{3}$ \\
$\lambda _{2}$ & $\lambda _{2}$ & $\lambda _{3}$ & $\lambda _{3}$ & $\lambda
_{3}$ \\
$\lambda _{1}$ & $\lambda _{1}$ & $\lambda _{2}$ & $\lambda _{3}$ & $\lambda
_{3}$ \\
$\lambda _{0}$ & $\lambda _{0}$ & $\lambda _{1}$ & $\lambda _{2}$ & $\lambda
_{3}$ \\ \hline
& $\lambda _{0}$ & $\lambda _{1}$ & $\lambda _{2}$ & $\lambda _{3}$%
\end{tabular}
\label{sml}
\end{equation}%
and where $\lambda _{3}\equiv 0_{S}$ is the zero element of the semigroup
such that $0_{S}\lambda _{\alpha }=0_{S}$. After considering a $0_{s}$%
-reduction, one finds a new algebra whose generators $\left\{ \tilde{J},%
\tilde{G}_{a},\tilde{H},\tilde{P}_{a},\tilde{Z},\tilde{Z}_{a}\tilde{S},%
\tilde{M},\tilde{T}\right\} $ are related to the Nappi-Witten ones as%
\begin{equation}
\begin{tabular}{ccc}
$\tilde{J}=\lambda _{0}\tilde{J}\,,$ & $\tilde{G}_{a}=\lambda _{0}\tilde{G}%
_{a}\,,$ & $\tilde{S}=\lambda _{0}\tilde{S}\,,$ \\
$\tilde{H}=\lambda _{1}\tilde{J}\,,\,$ & $\tilde{P}_{a}=\lambda _{1}\tilde{G}%
_{a}\,,$ & $\tilde{M}=\lambda _{1}\tilde{S}\,,$ \\
$\tilde{Z}=\lambda _{2}\tilde{J}\,,$ & $\tilde{Z}_{a}=\lambda _{2}\tilde{G}%
_{a}\,,$ & $\tilde{T}=\lambda _{2}\tilde{S}\,.$%
\end{tabular}%
\end{equation}%
Using the multiplication law of the semigroup (\ref{sml}) and the
Nappi-Witten commutators (\ref{NRLorentz}), one finds that the expanded
generators satisfy the following non-vanishing commutation relations%
\begin{eqnarray}
\left[ \tilde{G}_{a},\tilde{P}_{b}\right] &=&-\epsilon _{ab}\tilde{M}\,,%
\text{ \ \ \ \ \ \ }\left[ \tilde{G}_{a},\tilde{Z}_{b}\right] =-\epsilon
_{ab}\tilde{T}\,,\text{ \ }  \notag \\
\left[ \tilde{H},\tilde{G}_{a}\right] &=&\epsilon _{ab}\tilde{P}_{b}\,,\text{
\ \ \ \ \ \ \ \ \ \ }\left[ \tilde{J},\tilde{Z}_{a}\right] =\epsilon _{ab}%
\tilde{Z}_{b}\,,  \notag \\
\left[ \tilde{J},\tilde{P}_{a}\right] &=&\epsilon _{ab}\tilde{P}_{b}\,,\text{
\ \ \ \ \ \ \ \ \ }\left[ \tilde{H},\tilde{P}_{a}\right] =\epsilon _{ab}%
\tilde{Z}_{b}\,,  \label{MEBA} \\
\left[ \tilde{J},\tilde{G}_{a}\right] &=&\epsilon _{ab}\tilde{G}_{b}\,,\text{
\ \ \ \ \ \ \ \ }\left[ \tilde{P}_{a},\tilde{P}_{b}\right] =-\epsilon _{ab}%
\tilde{T}\,,  \notag \\
\left[ \tilde{G}_{a},\tilde{G}_{b}\right] &=&-\epsilon _{ab}\tilde{S}\,,%
\text{ \ \ \ \ \ \ \ \ }\left[ \tilde{Z},\tilde{G}_{a}\right] =\epsilon _{ab}%
\tilde{Z}_{b}\,.  \notag
\end{eqnarray}%
The algebra (\ref{MEBA}) correspond to the Maxwellian Exotic Bargmann
algebra and can be obtained as the NR contraction of [Maxwell] $\oplus
\,u(1)\oplus u(1)\oplus u\left( 1\right) $ algebra. Indeed, as was shown in
\cite{AFGHZ}, (\ref{MEBA}) can be obtained from the relativistic Maxwell
algebra generated by $\left\{ J\,_{A},P_{A},Z_{A}\right\} $ which satisfy%
\begin{eqnarray}
\left[ J_{A},J_{B}\right] &=&\epsilon _{ABC}J^{C}\,,\text{ \ \ \ \ }  \notag
\\
\left[ P_{A},P_{B}\right] &=&\epsilon _{ABC}Z^{C}\,,  \notag \\
\left[ J_{A},Z_{B}\right] &=&\epsilon _{ABC}Z^{C}\,,\text{ \ \ \ \ }
\label{relMaxwell} \\
\left[ J_{A},P_{B}\right] &=&\epsilon _{ABC}P^{C}\,,\text{ \ \ \ \ }  \notag
\end{eqnarray}%
and three extra $U\left( 1\right) $ generators $Y_{1}$, $Y_{2}$ and $Y_{3}$.
At the relativistic level, there has been a growing interest in exploring
(super)gravity models based on the Maxwell algebra and its generalizations
\cite{AKL, DKGS, AKL2, CPRS1, CPRS2, BGKL, AI, CR2, CFRS, CFR, PR, CCFRS,
Ravera, CPR, CRR, GKP, KSC, KC, Concha, SalgadoReb}. Interestingly a dual
version of the three-dimensional Maxwell algebra, known as Hietarinta
algebra \cite{Hietarinta}, can be obtained by interchanging the role of the
momentum generator $P_{A}$ with the Maxwell gravitational generator $Z_{A}$.
In this dual version, interesting results have been recently obtained in
\cite{BS, CS} in which a three-dimensional bi-gravity action has been
presented.

The contraction is defined through the following identifications:%
\begin{eqnarray}
J_{0} &=&\frac{\tilde{J}}{2}+\xi ^{2}\tilde{S}\,,\text{\ \ \ \ \ }J_{a}=\xi
\tilde{G}_{a}\,,\text{ \ \ \ \ \ \ }Y_{2}=\frac{\tilde{J}}{2}-\xi ^{2}\tilde{%
S}\,,  \notag \\
P_{0} &=&\frac{\tilde{H}}{2\xi }+\xi \tilde{M}\,,\text{ \ \ \ }P_{a}=\tilde{P%
}_{a}\,,\text{ \ \ \ \ \ \ \ }Y_{1}=\frac{\tilde{H}}{2\xi }-\xi \tilde{M}\,,
\\
Z_{0} &=&\frac{\tilde{Z}}{2\xi ^{2}}+\tilde{T}\,,\text{ \ \ \ \ }Z_{a}=\frac{%
\tilde{Z}_{a}}{\xi }\,,\text{ \ \ \ \ \ \ }Y_{3}=\frac{\tilde{Z}}{2\xi ^{2}}-%
\tilde{T}\,.  \notag
\end{eqnarray}%
which is the same identification used in the AdS-Lorentz case. The main
difference appears in the absence of the $\ell $ parameter in the
relativistic algebra (\ref{relMaxwell}).

It is worth it to mention that the MEB algebra\textit{\ }is obtained by
expanding the Nappi-Wittwn algebra using the same semigroup $S_{E}^{\left(
2\right) }$ used at the relativistic level \cite{CR1}. As we shall see, such
particular feature will appear not only for the MEB algebra. The same
behavior appears for infinite-dimensional expanded (super)algebras in which
the same expansion relation appearing for finite (super)algebras can be used
for infinite-dimensional (super)algebras \cite{CCRS, CCFR}.

The following diagram summarizes the NR limit as well as the $S$-expansion
applied at the relativistic and NR level:%
\begin{equation*}
\begin{tabular}{ccc}
\cline{1-1}\cline{3-3}
\multicolumn{1}{|c}{$%
\begin{array}{c}
\mathfrak{so}(2,1)\oplus u(1) \\
\left\{ J_{A},Y\right\}%
\end{array}%
$} & \multicolumn{1}{|c}{$\ \overset{S_{E}^{(2)}}{\longrightarrow }$ \ \ } &
\multicolumn{1}{|c|}{$%
\begin{array}{c}
\text{Maxwell}\oplus u(1)^{3} \\
\left\{ J_{A},P_{A},Z_{A},Y_{1},Y_{2},Y_{3}\right\}%
\end{array}%
$} \\ \cline{1-1}\cline{3-3}
&  &  \\
$\ \ \downarrow $ IW contraction &  & $\ \ \downarrow $ NR\text{ limit} \\
&  &  \\ \cline{1-1}\cline{3-3}
\multicolumn{1}{|c}{$%
\begin{array}{c}
\text{Nappi-Witten algebra} \\
\left\{ \tilde{J},\tilde{G}_{a},\tilde{S}\right\}%
\end{array}%
$} & \multicolumn{1}{|c}{\ \ \ $\overset{S_{E}^{(2)}}{\longrightarrow }$\ \
\ \ \ \ } & \multicolumn{1}{|c|}{$%
\begin{array}{c}
\text{Maxwellian Exotic Bargmann algebra} \\
\left\{ \tilde{J},\tilde{G}_{a},\tilde{H},\tilde{P}_{a},\tilde{Z},\tilde{Z}%
_{a},\tilde{S},\tilde{M},\tilde{T}\right\}%
\end{array}%
$} \\ \cline{1-1}\cline{3-3}
\end{tabular}%
\end{equation*}

\subsection{Enlargement of the Extended Bargmann Algebra}

Let us focus now in the NR version of the AdS-Lorentz algebra which is the
main topic of the present work. As we have seen, this algebra corresponds to
the contraction of the AdS-Lorentz algebra enlarged with three extra $U(1)$
generators, namely the NR limit of [AdS-Lorentz] $\oplus \,u(1)\oplus
u(1)\oplus u(1)$ algebra. An alternative way of deriving this NR algebra is
through the $S$-expansion.

As in the previous cases, we first consider the Nappi-Witten algebra (\ref%
{NRLorentz}) and a suitable semigroup. For our purpose, we choose the
Abelian semigroup $S_{\mathcal{M}}^{(2)}=\left\{ \lambda _{0},\lambda
_{1},\lambda _{2}\right\} $, whose elements satisfy%
\begin{equation}
\lambda _{\alpha }\lambda _{\beta }=\left\{
\begin{array}{lcl}
\lambda _{\alpha +\beta }\,\,\,\, & \mathrm{if}\,\,\,\,\alpha +\beta \leq 2
&  \\
\lambda _{\alpha +\beta -2}\,\,\, & \mathrm{if}\,\,\,\,\alpha +\beta >2 &
\end{array}%
\right.  \label{sm2}
\end{equation}%
The election of the semigroup is motivated by the fact that the AdS-Lorentz
algebra can be obtained as a $S_{\mathcal{M}}^{(2)}$-expansion of the
Lorentz algebra in three spacetime dimensions.

After considering the $S_{\mathcal{M}}^{(2)}$-expansion of the Nappi-Witten
algebra one finds an enlargement of the extended Bargmann algebra whose
generators $\left\{ \tilde{J},\tilde{G}_{a},\tilde{H},\tilde{P}_{a},\tilde{Z}%
,\tilde{Z}_{a},\tilde{S},\tilde{M},\tilde{T}\right\} $ satisfy the
commutation relations:%
\begin{eqnarray}
\left[ \tilde{G}_{a},\tilde{P}_{b}\right] &=&-\epsilon _{ab}\tilde{M}\,,%
\text{ \ \ \ \ \ \ }\left[ \tilde{G}_{a},\tilde{Z}_{b}\right] =-\epsilon
_{ab}\tilde{T}\,,\text{ \ \ \ \ \ \ }\left[ \tilde{H},\tilde{Z}_{a}\right] =%
\frac{1}{\ell ^{2}}\epsilon _{ab}\tilde{P}_{b}\,,  \notag \\
\left[ \tilde{H},\tilde{G}_{a}\right] &=&\epsilon _{ab}\tilde{P}_{b}\,,\text{
\ \ \ \ \ \ \ \ \ \ }\left[ \tilde{J},\tilde{Z}_{a}\right] =\epsilon _{ab}%
\tilde{Z}_{b}\,,\text{ \ \ \ \ \ \ \ }\left[ \tilde{Z}_{a},\tilde{Z}_{b}%
\right] =-\frac{1}{\ell ^{2}}\epsilon _{ab}\tilde{T}\,,  \notag \\
\left[ \tilde{J},\tilde{P}_{a}\right] &=&\epsilon _{ab}\tilde{P}_{b}\,,\text{
\ \ \ \ \ \ \ \ \ }\left[ \tilde{H},\tilde{P}_{a}\right] =\epsilon _{ab}%
\tilde{Z}_{b}\,,\text{ \ \ \ \ \ \ \ }\left[ \tilde{P}_{a},\tilde{Z}_{b}%
\right] =-\frac{1}{\ell ^{2}}\epsilon _{ab}\tilde{M}\,, \\
\left[ \tilde{J},\tilde{G}_{a}\right] &=&\epsilon _{ab}\tilde{G}_{b}\,,\text{
\ \ \ \ \ \ \ \ }\left[ \tilde{P}_{a},\tilde{P}_{b}\right] =-\epsilon _{ab}%
\tilde{T}\,,\text{ \ \ \ \ \ \ \ }\left[ \tilde{Z},\tilde{P}_{a}\right] =%
\frac{1}{\ell ^{2}}\epsilon _{ab}\tilde{P}_{b}\,,  \notag \\
\left[ \tilde{G}_{a},\tilde{G}_{b}\right] &=&-\epsilon _{ab}\tilde{S}\,,%
\text{ \ \ \ \ \ \ \ \ }\left[ \tilde{Z},\tilde{G}_{a}\right] =\epsilon _{ab}%
\tilde{Z}_{b}\,,\text{ \ \ \ \ \ \ \ \ }\left[ \tilde{Z},\tilde{Z}_{a}\right]
=\frac{1}{\ell ^{2}}\epsilon _{ab}\tilde{Z}_{b}\,.  \notag
\end{eqnarray}%
The explicit commutators appears by considering the multiplication law (\ref%
{sm2}) and the Nappi-Witten algebra (\ref{NRLorentz}). In particular, the
expanded generators are related to the Nappi-Witten ones as%
\begin{equation}
\begin{tabular}{lllllllllll}
$\tilde{J}$ & $=$ & $\lambda _{0}\tilde{J}$ & , & $\tilde{G}_{a}$ & $=$ & $%
\lambda _{0}\tilde{G}_{a}$ & , & $\tilde{S}$ & $=$ & $\lambda _{0}\tilde{S}%
\,,$ \\
$\ell \tilde{H}\,$ & $=$ & $\lambda _{1}\tilde{J}\,$ & , & $\ell \tilde{P}%
_{a}$ & $=$ & $\lambda _{1}\tilde{G}_{a}$ & , & $\ell \tilde{M}$ & $=$ & $%
\lambda _{1}\tilde{S}\,,$ \\
$\ell ^{2}\tilde{Z}$ & $=$ & $\lambda _{2}\tilde{J}\,$ & , & $\ell ^{2}%
\tilde{Z}_{a}$ & $=$ & $\lambda _{2}\tilde{G}_{a}\,$ & , & $\ell ^{2}\tilde{T%
}$ & $=$ & $\lambda _{2}\tilde{S}\,.$%
\end{tabular}%
\end{equation}%
It is interesting to note that the Nappi-Witten algebra is the smallest Lie
subalgebra of the EEB algebra containing rotations and boosts. The following
diagram summarizes the NR limits, the $S$-expansion applied at the
relativistic and NR level and the corresponding flat limits:%
\begin{equation*}
\begin{tabular}{ccccc}
\cline{1-1}\cline{3-3}\cline{5-5}
\multicolumn{1}{|c}{$%
\begin{array}{c}
\mathfrak{so}(2,1)\oplus u(1) \\
\left\{ J_{A},Y\right\}%
\end{array}%
$} & \multicolumn{1}{|c}{$\ \overset{S_{\mathcal{M}}^{(2)}}{\longrightarrow }
$ \ } & \multicolumn{1}{|c|}{$%
\begin{array}{c}
\text{\lbrack AdS-Lorentz]}\oplus u(1)^{3} \\
\left\{ J_{A},P_{A},Z_{A},Y_{1},Y_{2},Y_{3}\right\}%
\end{array}%
$} & \multicolumn{1}{|c}{$\overset{\ell \rightarrow \infty }{\longrightarrow
}$} & \multicolumn{1}{|c|}{[Maxwell]$\oplus u(1)^{3}$} \\
\cline{1-1}\cline{3-3}\cline{5-5}
&  &  &  &  \\
$\ \ \downarrow $ IW contraction &  & $\ \ \downarrow $ NR\text{ limit} &  &
$\downarrow $ NR limit \\
&  &  &  &  \\ \cline{1-1}\cline{3-3}\cline{5-5}\cline{5-5}
\multicolumn{1}{|c}{$%
\begin{array}{c}
\text{Nappi-Witten algebra} \\
\left\{ \tilde{J},\tilde{G}_{a},\tilde{S}\right\}%
\end{array}%
$} & \multicolumn{1}{|c}{\ \ \ $\overset{S_{\mathcal{M}}^{(2)}}{%
\longrightarrow }$\ \ \ \ \ } & \multicolumn{1}{|c|}{$%
\begin{array}{c}
\text{EEB algebra} \\
\left\{ \tilde{J},\tilde{G}_{a},\tilde{H},\tilde{P}_{a},\tilde{Z},\tilde{Z}%
_{a},\tilde{S},\tilde{M},\tilde{T}\right\}%
\end{array}%
$} & \multicolumn{1}{|c}{$\overset{\ell \rightarrow \infty }{\longrightarrow
}$} & \multicolumn{1}{|c|}{MEB$\text{ algebra}$} \\
\cline{1-1}\cline{3-3}\cline{5-5}\cline{5-5}
\end{tabular}%
\end{equation*}%
A particular advantage of the $S$-expansion procedure is that it allows us
to obtain the expanded invariant tensor in terms of the original one \cite%
{Sexp}. It is well known that the invariant tensor is a crucial ingredient
for the construction of a CS action. In particular, the non-vanishing
components of the invariant tensor of the Nappi-Witten algebra are given by%
\begin{eqnarray}
\left\langle \tilde{J}\tilde{S}\right\rangle &=&-1\,,\text{ \ \ \ \ \ \ \ \
\ \ \ \ \ \ \ \ \ \ \ \ \ \ \ \ \ \ }  \notag \\
\left\langle \tilde{G}_{a}\tilde{G}_{b}\right\rangle &=&\delta _{ab}\,,
\label{INVTL}
\end{eqnarray}%
Then, considering the definitions of the Theorem VII of \cite{Sexp}, we
recover the non-vanishing components of the EEB algebra given by (\ref{inv3a}%
)-(\ref{inv3b}). Let us note that the invariant tensor of the MEB algebra
can not only be derived as flat limit $\ell \rightarrow \infty $ of (\ref%
{inv3a})-(\ref{inv3b}) but can also be obtained from the Nappi-Witten
invariant tensor (\ref{INVTL}) considering $S_{E}^{\left( 2\right) }$ as the
relevant semigroup. It would be interesting to extend this methodology to
approach new supersymmetric extension of NR gravity theories.

\section{Discussion}

In this work we have presented the NR limit of the relativistic AdS-Lorentz
gravity theory. A particular $U\left( 1\right) $ enlargement of the
AdS-Lorentz algebra is considered in order to avoid infinity and degeneracy
difficulties. In particular, an enlargement of the extended Bargmann algebra
is presented by considering the NR contraction of the [AdS-Lorentz] $\oplus
\,u\left( 1\right) \oplus u\left( 1\right) \oplus u\left( 1\right) $
algebra. Such NR algebra allows to construct a proper finite NR CS action.
Interestingly, we have shown that the NR CS gravity theory presented here
reproduces the Maxwellian Exotic Bargmann gravity \cite{AFGHZ} in the limit $%
\ell \rightarrow \infty $. \ Furthermore, in such limit the NR CS gravity
action contains the Extended Bargmann gravity as a subcase.

We have also studied an alternative method to obtain the EEB and MEB algebra
using the $S$-expansion procedure. It is interesting to note that the $S$%
-expansion method not only provides us with consistent NR algebras but gives
us the appropriate central extension of the NR algebras in order to have
well defined non-degenerate bilinear form. In particular, at the
relativistic level, the same semigroup $S$ gives us the respective
relativistic algebras with extra Abelian generators whose presence assures a
finite Lagrangian in the NR limit. Let us note that further generalizations
of the Galilean algebra have also been obtained through the $S$-expansion
method in \cite{GRSS}. \ On the other hand interesting works have recently
appeared in the literature where the Lie algebra expansion method using the
Maurer-Cartan equations \cite{AIPV}, which is the first report introducing in a general procedure the Lie algebra expansions, has been used to obtain several
(super)algebras for NR\ (super)gravity \cite{BIOR, AGI}. It would be worth studying the possibility to obtain the EEB algebra and the MEB one by applying the Lie algebra expansion of \cite{AIPV} to the relativistic AdS-Lorentz and Maxwell algebra, respectively.

It would be interesting to extend the results obtained here at the
supersymmetric level. In particular, the Maxwell and AdS-Lorentz
supergravity in the CS formalism have been explored recently in \cite{CPR,
Concha}. It would be worth it to explore the extra bosonic field required at
the relativistic level to formulate a proper finite NR supergravity action
[work in progress].

Another aspect that deserves further investigation is the ultrarelativistic
limit of the AdS-Lorentz theory [work in progress]. One could obtain the
complete cube, analogously to the ones presented in \cite{BLL, MPS},
describing the ultrarelativistic, non-relativistic and flat limits for the
AdS-Lorentz symmetry. In particular, one could expect to find an Carroll
version of the AdS-Lorentz algebra.

\section*{Acknowledgment}

The authors would like to thank D. Hidalgo and J. Matulich for valuable
comments. This work was supported by the Chilean FONDECYT Projects N$^{\circ
}$3170437 (P.C.) and N$^{\circ }$3170438 (E.R.).

\end{document}